\documentclass[onecolumn,11pt,aps,prd,superscriptaddress,nofootinbib,longbibliography]{revtex4-2}
\usepackage{amsmath,amssymb}
\usepackage[colorlinks,linkcolor=red,anchorcolor=blue,citecolor=cyan]{hyperref} 
\usepackage{graphicx}
\usepackage{adjustbox}
\usepackage{tikz}
\usetikzlibrary{shadings}
\usetikzlibrary{arrows.meta}
\usetikzlibrary{calc}
\usepackage{colortbl}
\usepackage{xcolor}
\usepackage{picinpar}
\usepackage{float}
\usepackage{wrapfig}

\usepackage{slashed}
\usepackage{ulem}
\usepackage{bm}
\usepackage{mathrsfs}
\usepackage{array,cases}
\usepackage{comment}
\usepackage{multirow,paracol}

\usepackage[final]{changes}
\usepackage{lipsum}
\usepackage{amsmath,mathtools,amssymb}
\usepackage{bm,extarrows,ulem,cancel,slashed}
\usepackage{mathrsfs}
\usepackage{esint}
\usepackage{geometry,graphicx,color}
\usepackage[all,pdf]{xy}
\usepackage{pstricks,pstricks-add}
\usepackage{hyperref}
\hypersetup{pdfborder=001,colorlinks=true, linkcolor=black}
\usepackage{listings}

\newcommand{\be}{\begin{equation}}
\newcommand{\ee}{\end{equation}}
\newcommand{\bea}{\begin{eqnarray}}
\newcommand{\eea}{\end{eqnarray}}
\newcommand{\ba}{\begin{array}}
\newcommand{\ea}{\end{array}}
\newcommand{\bs}{\be\begin{split}}
\newcommand{\es}{\end{split}}

\newcommand{\mi}{\mathrm{i}}
\newcommand{\me}{\mathrm{e}}

\renewcommand{\1}{\left}
\renewcommand{\2}{\right}
\newcommand{\ma}{\mathcal}
\newcommand{\ri}{\rightarrow}

\newcommand{\rom}[1]{\uppercase\expandafter{\romannumeral #1}}

\newcommand{\m}{\mu}
\newcommand{\n}{\nu}

\newcommand{\ep}{\epsilon}

\newcommand{\bes}{\be\1\{\begin{split}}

\def\roughly#1{\mathrel{\raise.3ex\hbox{$#1$\kern-.75em%
\lower1ex\hbox{$\sim$}}}}

\def\({\left(}
\def\){\right)}
\def\[{\left[}
\def\]{\right]}
\def\<{\langle}
\def\>{\rangle}

\def\l{{\lambda}}

\def\o{{\omega}}

\def\m{{\mu}}
\def\n{{\nu}}

\newcommand{\diff}{\mathrm{d}}

\allowdisplaybreaks

\begin{document}
	
	
	\author{Zhiming Shuai}
	\email{202410188403@mail.scut.edu.cn}
	\affiliation{School of Physics and Optoelectronics, South China University of Technology, Guangzhou 510641, China}
	
	\author{Xiangdong Zhang}
	\email[Corresponding author: ]{scxdzhang@scut.edu.cn}
	\affiliation{School of Physics and Optoelectronics, South China University of Technology, Guangzhou 510641, China}
	
	\author{Gui-Rong Liang}
	\email[Corresponding author: ]{bluelgr@sina.com}
	\affiliation{School of Materials Science and Physics, China University of Mining and Technology, Xuzhou 221116, China}

	
	\title{Scalar superradiance in the charged black-bounce spacetimes}

	\begin{abstract}
		We numerically investigate the superradiant amplification effect of a charged scalar field in the scattering experiment and the black hole bomb model in a charged black-bounce spacetime. Due to the shallowing effect on the effective potential by the introduced quantum parameter $\l$, superradiance in both the above cases is verified to be weakened. In a scattering experiment, the quantum parameter and the field mass suppress the amplification in all frequency ranges, while the black hole and field charge suppress the amplification at low frequencies but enhance it at high frequencies. In a Type I black hole bomb model, where the reflective mirror is placed outside the ergo-region, we find a new distinct eigen-mode for the scalar field evolution in a high $\l$ value, which is however absent in the light-particle sub-case of the Type II black hole bomb where the mirror is set inside the ergo-region. Moreover, we investigate the heavy field mass scenario in a Type II black hole bomb, in which an additional eigenfrequency emerges at the same high $\l$ value but is not superradiantly amplified.
	\end{abstract}
	\maketitle
	
	\newpage
	\section{Introduction}
	The phenomenon of superradiance exhibits wave amplification of fields in a thermodynamic system, with the cost of dissipation in a ``viscous medium".
	The first example was given by the so-called ``Klein paradox" \cite{Klein:1929aa, Sauter:1931aa}, in which charged bosons get energy enhancement when scattered by a potential barrier where particle-antiparticle pair creations are excited \cite{Brito_2020}; the antiparticles with negative energy were injected into the barrier and particles with positive energy join into the original particle beam and are reflected. Energy conservation is kept and necessary during the process. Fermions do not possess such amplification due to the positive definiteness of its energy current density, 
	in the quantum field point of view, 
	it is Pauli exclusion principle that limits numbers of pair productions \cite{MANOGUE1988261} to form a macroscopic negative current. 
	In curved spacetime, superradiance can also occur in vacuum \cite{Brito_2020};
	especially in black hole (BH) physics, the existence of the event horizon renders the BH behave as a ``viscous one-way membrane", such that an intrinsic dissipative region is provided
	in the form of an ergoregion, where timelike particles can have negative energies with respect to infinite observers. In the Penrose thought experiment \cite{Penrose:1969pc}, particle splitting takes place in a similar manner, however in a classical sense, with that of pair creation in the Klein paradox, such that energy of BHs is extracted, which happens both in rotation and charged static background. The wave analogue comes parallel in the low-frequency regime, followed by quantitative analysis in pioneering works in Kerr  \cite{Teukolsky:1974yv,PRESS:1972aa} and Reissner-Nordstr\"om (RN) \cite{Denardo:1973pyo} spacetimes.
	The ergoregion is intrinsic in rotating case, but an effective description in charged static case where it includes both information of the curvature and properties of the field, e.g., mass and charge, and may even cover the whole exterior of the black hole \cite{Laurent_Di_Menza_2015}.
	
	Superradiant instabilities occur in the sense that the process can be carried repeatedly, when a reflective mechanism is provided in a distant region to bounce the amplified wave back again. In the rotating case, the mass of the bosonic field itself serves as a natural mirror to bound the condensing cloud, while asymptotically-flat charged BHs were shown to be stable against massive charged scalar perturbations \cite{HOD2012505,HOD20131489}, hence a reflective boundary must be imposed in a toy model, to make a so-called ``black hole bomb" \cite{PRESS:1972aa}. 
	In contrast to the astronomical relevance in the rotating case \cite{Endlich_2017,Dolan_2013,Jia:2023see,PhysRevLett.119.041101,PhysRevD.76.084001,HUANG2019135026,Franzin_2021,Khodadi_2021,Kobayashi_2010,LI2023137902,Chen:2022nbb,Roy:2021uye,Jha:2022tdl,Khodadi_2022,Khodadi_2021_2}, the increasing investigations on superradiance in charged static system \cite{Mascher_2022,Benone_2014,Benone_2016,Zhu_2014,Konoplya:2014lha,Di_Menza_2020,PhysRevLett.116.141101,Degollado:2013aa} make it more interesting from a conceptual point of view: Its remarkably short instability time scale
	makes it an ideal testbed for fully nonlinear studies of black hole instabilities within cavities, in \cite{Sanchis-Gual:2016ab}, the researchers employ an asymptotically flat, spherically symmetric spacetime with a boundary at a fixed coordinate distance where the charged scalar field adheres to Dirichlet conditions. This framework permits tracking the linearized instability into the nonlinear regime, with substantial evidence indicating a final entropically preferred state where a black hole is surrounded by a charged condensate \cite{Dolan:2015aa}. 
	Moreover, the occurrence of superradiance in static charged black holes plays a vital role in enabling diverse applications within the gauge/gravity duality framework. Notable examples include the mechanism of spontaneous symmetry breaking in the vicinity of a Reissner-Nordström anti-de Sitter (RN-AdS) black hole \cite{Gubser:2008aa}, as well as associated developments in holographic superconductor models \cite{Hartnoll:2008aa}.
	Research on charged black hole bombs has progressed through both analytical treatment \cite{Hod_2013,Li:2015ab}, and numerical approaches in which the behaviors were examined at the linear field amplitude level  in both frequency and time-domain \cite{Degollado:2013aa,Degollado:2014aa,Dias_2018}. 
	Generalizations were carried for a charged massless scalar field in the spacetime of a charged stringy black hole with mirror-like boundary conditions \cite{Li:2015aa,LI2015520}, and in the context of massive charged scalar fields on a Kerr-Newman background \cite{10.1143/PTP.112.983}, which confirms even in the rotation-dominated regime
	the instability growth rate depends on the coupling between the field charge and the black hole charge.

	On the other hand, it is widely believed that quantum gravity effects would remove the curvature singularity of a black hole and result in a regular/non-singular black hole \cite{Ashtekar:2023cod}. However, since a complete theory of quantum gravity is not yet well established, investigations on various non-singular black holes at the phenomenological level were extensively carried out. Although these models are usually not any solutions to the vacuum Einstein equation, one can fix the logic by introducing some exotic matter field on the right hand side of the equation.
	Regular black holes can be traced back to the well known Bardeen black hole \cite{Bardeen}, Hayward black hole \cite{Hayward} as well as other notable regular black hole models \cite{Frolov,Ay_n_Beato_1998,Fan_2016,Ghosh_2021,Barrientos_2022}. The properties of regular black holes were studied
	with the purpose of revealing the internal structure of black holes and capturing the signals due to the possible quantum gravity effect. 
	In recent years, investigations on the superradiant effect and black hole bomb instability of different types of regular black holes were carried increasingly
	\cite{dolan_charged,paula_absorption,Yang_2023,Zhan_2024gvi,Li_zhen,Franzin}, 
	which provided interesting insights in understanding the interplay between fundamental and phenomenological aspects.
	
	
	In this paper, we investigate the superradiant effects in
	a spherically symmetric charged black-bounce spacetime \cite{Franzin_bbrn,GUO2022115938},
	which allows a special class of regular black holes and 
	wormhole solutions in a smooth parameter tuning. 
	The metric could be achieved by a radial coordinate transformation 
	$r\to \sqrt{r^2+\lambda^2}$, with $\lambda$  a length scale parameter representing quantum regularization, to deform the classical Reissner-Nordstr\"om (RN) metric
	to the black-bounce-RN geometry \cite{Simpson_2019,Simpson_vaidya,Lobo_2020,Lobo_2021,Mazza_2021}. 
	The introduced parameter is typically associated with the Planck length, which smooths the central singularity while preserving asymptotic RN behavior. 
	The model provides a phenomenological framework to probe how quantum gravitational effects alter the dynamics of superradiant processes and black hole bombs, and allows for a controlled comparison with classical results by tuning $\lambda$. The primary focus of this work is to analyze how this length scale parameter influences two interconnected phenomena, the efficiency of superradiant energy extraction from the black hole, and the instabilities of black hole bomb.
	
	The remainder of the paper is organized as follows. 
	In section \ref{sec1}, we will briefly review basic properties of the black-bounce-RN spacetime, as well as the dynamics of a charged scalar field propagating on this background. By the Regge-Wheeler analysis, we extract the effective potential, and investigate how it is influenced by different choices of the quantum parameter. Through physical inspections, we predict the tendency of how the parameter $\l$ would affect the superradiant amplification, which is to be verified and confirmed in following contexts. Then we numerically explore the influence of each parameter in the metric on the amplification factor in a scattering experiment, and observe the dependence in each frequency band. 
	In section \ref{sec2}, we numerically study the superradiant amplification of the scalar field in two types of black hole bomb models, where the reflective mirror described by a Dirichlet boundary condition is put outside or inside the ergo-region. To mimic the purely ingoing boundary condition, the perfectly matched layer (PML) \cite{Antoine18082017,absobing_bjorn,APPELO2007531} method is employed and tested. In solving the differential equation in the time domain, the finite difference method combined with the method of lines are used, and the eigenvalues are extracted from the wave amplitude evolution by a discrete Fourier transform. To gain the energy flux, we invoke a type of constructed conserved energy current out of the energy-momentum tensor. In a Type II black hole bomb model, we further investigate the heavy field mass scenario where the field is in a confined configuration. Results are shown mainly in the form of diagrams, and we exhibit findings in the adjoint contexts. In section \ref{sec3}, we conclude and re-organize the logic presented in the main text, and outlook on some possible extensions.
	Throughout the paper, we adopt the $(-,+,+,+)$ signature in the metric and the natural units $c=G=\hbar=1$.
	
	\newpage
	\section{Dynamics of charged scalar fields in black-bounce-Reissner-Nordstr\"om spacetimes}\label{sec1}
	

	In this section, we will briefly review the black-bounce-Reissner-Nordstr\"om spacetimes \cite{Franzin_bbrn}, inspired by the uncharged black-bounce spacetime \cite{Simpson_2019}, which is a type of black hole without singularity. Performing the regularizing replacement $ r\rightarrow \sqrt{r^2+\lambda^2} $ in the RN metric, with $\lambda$ 
	a length scale 
	parameter typically associated with the Planck length, we obtain the following line element
	\begin{equation}
		\1\{\begin{aligned}
			&\diff s^2=-f(r)~\diff t^2+f(r)^{-1}~\diff r^2+(r^2+\lambda^2)~\diff\Omega^2,\\
			& f(r)=1-\frac{2M}{\sqrt{r^2+\lambda^2}}+\frac{Q^2}{r^2+\lambda^2},
		\end{aligned}\2.
	\end{equation}
	with $M$ and $Q$ the BH mass and electric charge. Note that the replacement extends the natural domain of the radial coordinate $r$ from $[0, +\infty]$ to $[-\infty, +\infty]$, and asymptotic flatness is preserved.
	By computing the Kretschmann scalar and curvatures, one verifies that this replacement renders the geometry globally regular, rather than simply making a coordinate transformation \cite{Franzin_bbrn}.
	The metric is controlled precisely in a minimal manner to describe regular BHs and traversable wormholes smoothly, as we can check that when $M\ri 0$ and $Q\ri 0$ it recovers the standard Morris-Thorne wormhole \cite{wormhole1,PhysRevLett.61.1446,Boonserm:2018orb,Franzin_bbrn}.
	The horizons can be obtained by $f(r)=0$ which gives
	\begin{equation}
		r_\pm=\sqrt{\left(M\pm\sqrt{M^2-Q^2}\right)^2-\lambda^2},
	\end{equation}
	where
	the existence of horizons is guaranteed by a range of parameter choices satisfying $|Q|<M$ and 
	$ |\lambda|<M+\sqrt{M^2-Q^2} $.
	
	Though the purpose of the metric is due to quantum gravity effects, we can also understand it in a way that the metric
	is described by general relativity, with the determination of the associated non-trivial matter/energy content is accomplished by the Einstein equation, with which we could fix the geometro-dynamics.
	It is verified that outside the outer horizon, the radial null energy condition (NEC) is violated, hence the weak, strong and dominant energy conditions are all violated \cite{Simpson_2019}.
	With the help of Einstein equation, non-zero components of the energy-momentum tensor can be computed.
	The explicit form of the tensor can be decomposed into two parts, in which one is the
	usual Maxwell stress-energy tensor, and the other is interpreted as the ``charged dust" \cite{Franzin_bbrn}, with its density involving both the bounce parameter $\lambda$ and the total charge $Q$. 
	
	The electric field strength $E$ is obtained as
	\be
	E=
	\frac{Qr}{(r^2+\l^2)^{3/2}}=E_\text{RN}\[\frac{r^3}{(r^2+\l^2)^{3/2}}\],
	\ee
	with $E_\text{RN}$ the electric field strength of the RN black hole. One can check it by the Gauss theorem, that the total charge $Q$ is not totally enclosed in the event horizon due to the absence of the inverse square law of the electric field; actually in the ``charged dust" scenario, the BH charge distribution is somewhat diluted, and diffused in the whole spacetime, 
	including outside the horizon, 
	although its density decays very fast as the inverse sixth power of $r$ \cite{Franzin_bbrn}.
	The electromagnetic potential is extracted by integrating the above field and setting the integration constant to zero to ensure the asymptotic flatness, thus we have
	\begin{equation}\label{sec1:electromagnetic potential}
		A_{\mu}=-\frac{Q}{\sqrt{r^2+\lambda^2}}(1,0,0,0).
	\end{equation}
	which is simply 
	a replacement $r\rightarrow \sqrt{r^2+\lambda^2}$ in the RN electromagnetic potential. As we can see, at a fixed radial coordinate, this spacetime possesses weaker electrical potential compared to that of RN case.
	
	\subsection{Regge-Wheeler analysis and the effective potential}\label{subsec1}
	
	The Lagrangian for a charged complex scalar field is given by
	\begin{equation}\label{sec3:Lagrangian}
		\mathcal{L}=
		(D^\nu\Phi)^*D_\nu\Phi
		+\mu^2\Phi^*\Phi
	\end{equation}
	with $D_\nu=\nabla_\nu+\mi qA_\nu$ the gauge-covariant derivative chosen to satisfy  $[D_\nu, D_\rho]=\mi q F_{\nu\rho}$,
	and $\mu$ and $q$ the mass and charge of the scalar field respectively. 
	The corresponding Klein-Gordon equation is
	\begin{equation}\label{KG equation}
		D_\nu D^\nu\Phi-\mu^2\Phi=0.
	\end{equation}
	In the time domain, the scalar field admits separable solutions of the form
	\begin{equation}\label{sec3:seperation of variables}
		\Phi=\sum_{lm}\frac{\psi(t,r)}{\sqrt{r^2+\lambda^2}}Y_{lm}(\theta,\phi),
	\end{equation}
	where $ Y_{lm} $ is spherical harmonic function. The temporal part can be further converted to frequency domain by a Fourier transform $\psi(t,r)=\int \diff \o ~\psi_{\o l}(r) \me^{-\mi \o t}$.
	Substituting these decomposition of componenets 
	into Klein-Gordon equation \eqref{KG equation}, and invoking the tortoise coordinate $x$ as $\diff x/\diff r={1}/{f(r)}$, we can get the 
	differential equation for the radial part in the time domain,
	\begin{equation}\label{sec3:time-domain eom}
		\partial_x^2\psi-\partial_t^2\psi-2\mi qA_0\partial_t\psi-\[W_\text{eff}- (qA_0)^2\]\psi=0,
	\end{equation}
	and in the frequency domain,
	\begin{equation}\label{sec2:radial equation}
		\partial_x^2\psi_{\omega l}+\left[ \1(\omega-qA_0\2)^2-W_\text{eff} 
		\right] \psi_{\omega l}=0,
	\end{equation}
	with the effective potential $W_{\text{eff}}$ explicitly expressed as
	\begin{equation}\label{sec2:effective potential}
		\begin{aligned}
			W_{\text{eff}}=&f(r)\left[ \frac{l(l+1)}{r^2+\lambda^2}+\mu^2 +f(r)\frac{\lambda^2}{(r^2+\lambda^2)^2} 
			+f^\prime (r)\frac{r}{r^2+\lambda^2} \right]
		\end{aligned}
	\end{equation}
	which is a generalization of Regge-Wheelar analysis in \cite{Boonserm:2013dua,Boonserm:2018orb} with the field charge and mass, in the metric with BH charge and quantum parameter. One can recover the usual RN effective potential by setting $\l\ri 0$. Since the effective potential has an intimate relation with the definition of ergo-region \cite{Brito_2020}, which further affects the superradiance magnitude, it is worthwhile to 
	investigate how each parameter
	 influences the potential. We adopt the definition in \cite{Laurent_Di_Menza_2015}: the effective ergo-region is where the quantity below is negative:
	\begin{equation}
		V_\text{eff}=W_\text{eff}- (qA_0)^2.
	\end{equation}
	This definition singles out terms containing $\o$ in Eq.~\eqref{sec2:radial equation} and is convenient in time-domain analysis which we will carry out later. 
	\begin{figure}[htbp!]	
		\centering
		\includegraphics[scale=0.35]{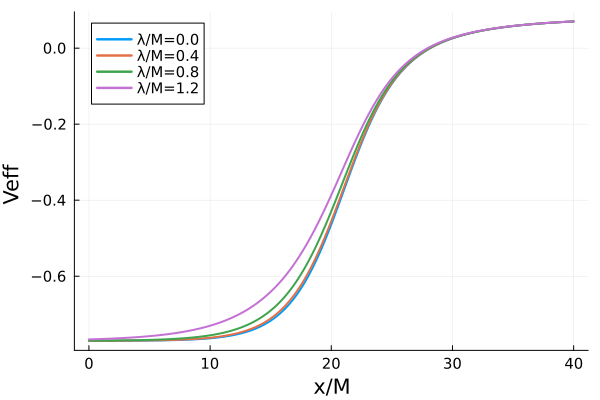}
		\includegraphics[scale=0.35]{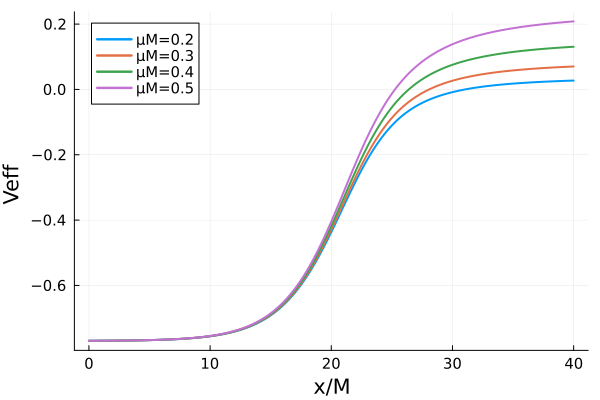}
		\includegraphics[scale=0.35]{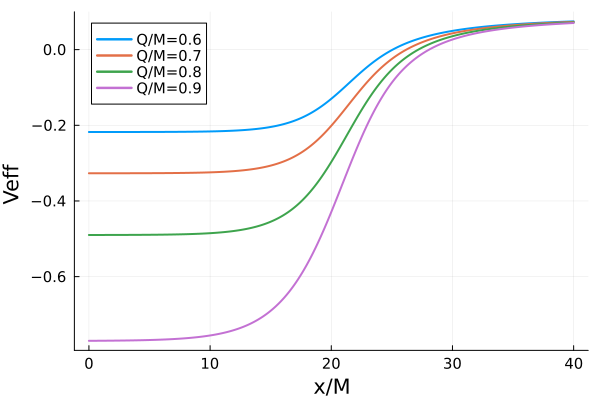}
		\includegraphics[scale=0.35]{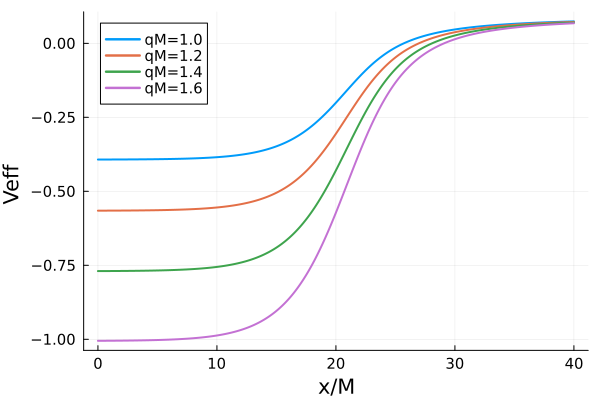}
		\caption{The spatial distribution of the effective potential is shown under different parameters. The standard parameters are set as $\lambda/M=0.8,\mu M =0.3,qM=1.4,Q/M=0.9$ and $l=0$. The four plots illustrate variations in $\lambda/M,\mu M,Q/M$ and $qM$, respectively.}
		\label{sec3:potential}
	\end{figure}
	In Fig. \ref{sec3:potential}, we show the effective potential for different parameters. The standard parameters are chosen as  $\lambda/M=0.8,\mu M =0.3,qM=1.4,Q/M=0.9$ and $l=0$ of black hole and scalar field. We then individually vary the parameters $ \lambda/M $, $ \mu M $, $Q/M $, and $ qM $ in the subsequent analyses, respectively. The roles of these four parameters can be categorized into three types. For the first type, corresponding to $\lambda/M$, the effective potential values at both sides remain identical with different $\lambda$ and it undergoes a drastic change around $x/M=20$, while the variation of the effective potential gradually becomes smoother as $\lambda$ grows larger. As the potential well becomes ``shallower", the extent of negativity a particle's energy can reach is bounded, thus we could expect, the superradiant ability in the regular background is weakened, as will be confirmed in both the 
	scattering experiment and 
	the BH bomb scenario below. The second category corresponds to $ \mu M $. Modifying this parameter affects the value of the effective potential far from the black hole event horizon, and as it increases,
	 the range where the potential takes a negative value becomes ``narrower",
	therefore we expect the amplification factor to gradually decrease with increasing $ \mu M $. The final category includes $ qM $ and $ Q/M$,
	 as the charges increase, the effective potential near the BH horizon continuously decreases. Due to this ``deepening effect", 
	  particles are expected to be able to carry away more energy and give rise to a larger superradiant amplification.

		\subsection{The superradiance amplification factor in a scattering experiment}
		It's useful to consider the asymptotic behaviors of the differential equation \eqref{sec2:radial equation} to establish a connection between the wave flux at the event horizon and spatial infinity.
		We assume at infinity there's an incident and reflected wave, nevertheless at the horizon there's only purely ingoing wave, without any quantum-induced reflective part as in \cite{Guo:2021xao, Jia:2023see, Luo:2024gqo}, thus the solution
		behaves as
		\begin{equation}\label{sec2:asymptotic solution}
			\psi_{\omega l}=\left\{\begin{aligned}
				\mathcal{I}\me^{-\mi k_{\infty} x}+\mathcal{R} \me ^{\mi k_{\infty} x},& \quad r\rightarrow\infty\\
				\mathcal{T}\me^{-\mi k_{+} x},&\quad r\rightarrow r_+
			\end{aligned}\right. ,
		\end{equation}
		with $k_{\infty}=\sqrt{W_\text{eff}}=\sqrt{\omega^2-\mu^2}$ and $ k_{+}=\omega-qA_0(r_+) $ the square root of the asymptotic effective potential, and $\mathcal{R},\ \mathcal{T}$ the reflection and transmission coefficient respectively. Moreover, due to the conservation of flux (Wronskian), we have the relation
		\begin{equation}\label{sec2:reflection and transmission}
			|\mathcal{R}|^2+\frac{k_+}{k_{\infty}}|\mathcal{T}|^2=|\mathcal{I}|^2.
		\end{equation}
		It is seen that when $ k_+<0 $, i.e., $ \omega<qA_0(r_+) $, the reflection coefficient exceeds the incident coefficient $|\mathcal{R}|^2>|\mathcal{I}|^2$, which indicates the scalar field is amplified through superradiance and the energy can be extracted from the black hole.
		A dimensionless 
		amplification factor \cite{Brito_2020} is defined,
		\begin{equation}
			Z_{l}=\left|\frac{\mathcal{R}}{\mathcal{I}}\right|^2-1,
		\end{equation}
		to measure the effect of superradiance in different choices of parameters.
		Below we compute it by the numerical methodology  commonly employed in the literature \cite{Brito_2020}.
		The key is to solve the governing equation \eqref{sec2:radial equation}, in which we impose the purely ingoing boundary condition \eqref{sec2:asymptotic solution} at a radial coordinate approaching the event horizon, specifically at $ r=r_+ (1+\epsilon) $ with $ \epsilon = 10^{-5} $, then we numerically integrate the radial equation  \eqref{sec2:radial equation} starting from the vicinity of the event horizon to a region far from the black hole to determine the solution at $ r_\infty=1000r_+ $, where the numerical solution has stabilized. 
		The reflection and incident coefficients are identified by comparing it with the asymptotic form derived for $ r\rightarrow \infty $ in equation \eqref{sec2:asymptotic solution}, thus the amplification factor is obtained.
		
		It is natural to expect that parameters in both the background and the field could influence this factor,  
		below we focus on the $l=0$ mode, and specifically choose $\lambda/M=0.8, Q/M=0.9, qM= 1.4, \mu M=0.3$ as our standard value, where the BH mass is used to make other quantities dimensionless. We vary each parameter in the following figures
		and present our numerical result 
		of the amplification factor as a function of frequency $\omega$ in different parameter configurations.

\begin{figure}[h]
			\centering
			\includegraphics[scale=0.35]{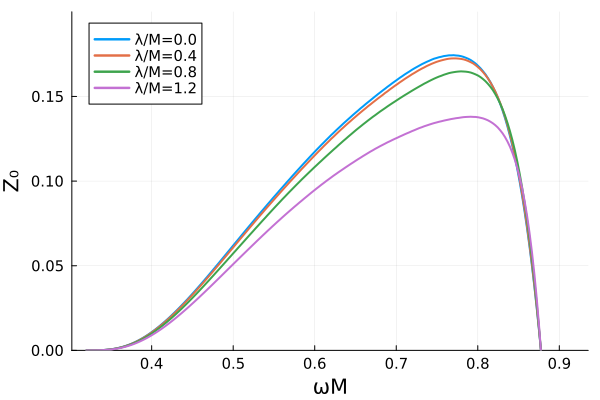}
			\includegraphics[scale=0.35]{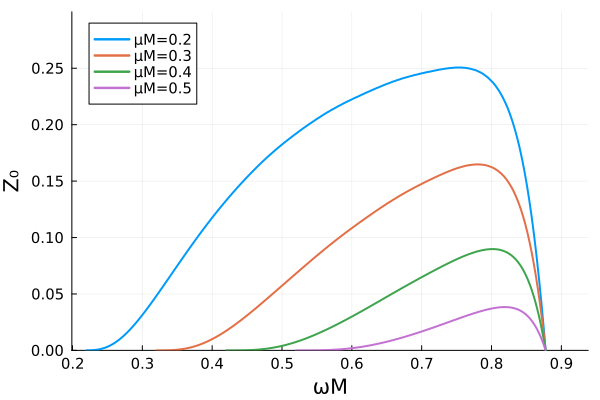}
			\caption{Amplification factor of the scalar field as a function of the frequency with dependence on different choices of $\lambda/M$ (left) and on different field mass $\m M$ (right). Curves in blue, orange, green, purple color in the left panel denote cases of $\lambda/M=0, 0.4, 0.8, 1.2$ respectively with $\m M=0.3$ fixed, and in the right panel denote $\m M=0.2, 0.3, 0.4, 0.5$ respectively with $\l/M=0.8$ fixed. Other parameters are chosen as the standard values, $Q/M=0.9, qM=1.4, l=0$. The point where curves meet the horizontal axis gives the critical superradiant frequency, which is $\omega_cM=0.8775$ in these cases. 
			}
			\label{sec2:fig4}
		\end{figure}

		Firstly and importantly, we check the parametric dependence on $\lambda/M$, which is the distinct feature of the regular metric. In the left panel of Fig.\ref{sec2:fig4},
		it is seen that with the increase of the quantum parameter $\lambda/M$, the magnitude of amplification factor decreases in a wide range of frequencies, as expected in Subsection~\ref{subsec1} due to the weakening effect on the potential well. More notably, we observe that the reduction rate of the amplification factor exhibits progressive acceleration with successive increasing values of the dimensionless quantum parameter $\lambda/M$. Due to the similarity of the overall trend, we also list the dependence on different choices of the scalar field mass $\m M$ on the right panel of Fig.\ref{sec2:fig4}, and we can find that the amplification factor gradually decreases, in all ranges of $\o$, with the increasing of $\m M$. But note that numerical solutions are obtained far away from the black hole, the condition $ \omega>\mu $ must hold, thus we can observe that all amplified frequencies are greater than the scalar field mass. In both figures, the critical superradiant frequency is $\omega_cM=0.8775$, which is not affected by $\l/M$ and $\m M$.
		
\begin{figure}[h]
			\centering
			\includegraphics[scale=0.35]{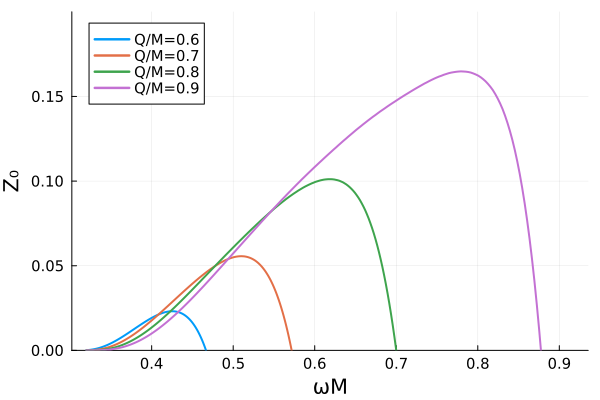}
			\includegraphics[scale=0.35]{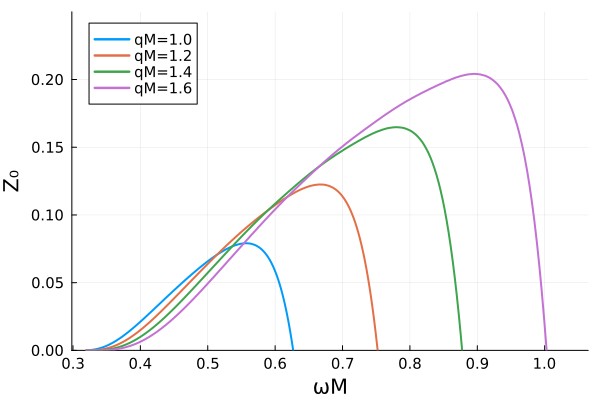}
			\caption{Amplification factor of the scalar field as a function of the frequency with dependence on different choices of $Q/M$ (left) and on $qM$ (right). Curves in blue, orange, green, purple color in the left panel denote cases of $Q/M=0.6, 0.7, 0.8, 0.9$ respectively with $q M=1.4$ fixed, and in the right panel denote $qM=1, 1.2, 1.4, 1.6$ respectively with $Q/M=0.9$ fixed. Other parameters are chosen as the standard values, $\l/M=0.8, \m M=0.3, l=0$. }
			\label{dep_charge}
		\end{figure}

		Another group of parameters are the BH and the field charge, which do affect the critical superradiant frequency as $\o=q A_0$, as the increase of the charges leads to substantial broadening of the cutoff frequency range.
		We present the dependence of the amplification factor on $Q/M$ and $qM$ on the left and right panels of Fig.\ref{dep_charge}.
		We can find a similar trend that with the increases of the BH or field charge, the amplification factor decreases in the low-frequency regime, but increases in the high-frequency regime. 
		
		In the above analysis, the quasi-stationary approximation is implied, i.e., fluctuations of order $\ma O(\ep)$ in the scalar field in the background induce changes in the spacetime geometry of order $\ma O(\ep^2)$, due to the fact that the stress-energy tensor is quadratic in the fields, and therefore to leading order can be studied on a fixed geometry \cite{Brito_2020}. 
		Up to now, we have discussed the parameter dependence of a regular BH to the superradiant scattering amplification factor. The main ingredient is coded in the effective potential. Later we will see this also holds in the case of a BH bomb scenario.

		\section{Black hole bombs of Type I and II in black-bounce-Reissner-Nordstr\"om spacetimes}\label{sec2}	
		As we have mentioned in the introduction, a scenario where a black hole is enclosed by reflective mirrors, the superradiant mechanism can induce amplification of scalar fields confined between the event horizon and the mirror boundary. 
		The field mass in the Kerr metric could serve as a natural mirror, while in RN due to the stability of the spacetime only theoretical reflective boundary condition can be used to confine the field. This energy accumulation in the scalar field may eventually break the mirror, leading to the black hole bomb phenomenon \cite{PRESS:1972aa}. This section aims to numerically investigate black hole bombs on the black-bounce-RN black hole.
		The definition of the conserved energy should be firstly given.
		
		\subsection{The energy current for the charged scalar field}
		The symmetric energy-momentum tensor $T_{\mu\nu}$ for the charged scalar field corresponding to 
		the Lagrangian \eqref{sec3:Lagrangian} is given by
		\begin{equation}
			\begin{aligned}
				T_{\mu\nu}=D_{(\mu}\Phi(D_{\nu)}\Phi)^*-g_{\mu\nu}\mathcal{L}.
			\end{aligned}
		\end{equation}
		which is not divergence-less due to the exclusion of the background electro-magnetic term in the Lagrangian, physically the coupling back is ignored \cite{Laurent_Di_Menza_2015,Di_Menza_2020}. Hence the energy current $J_\n\equiv T_{\n 0}$ is not conserved and the covariant derivative is explicitly given as
		\begin{equation}\label{sec3:covariant derivative of total current}
			\nabla_\nu J^\nu=qF_{\nu 0} (\phi_2\nabla^\nu\phi_1-\phi_1\nabla^\nu\phi_2),
		\end{equation}
		where $\phi_1$ and $\phi_2$ represent the real and imaginary parts of the scalar field.  
		Via eliminating the electric potential from the energy current we can construct a conserved energy current, which encodes the intrinsic energy of the scalar field. The conserved energy current reads
		\begin{equation}\label{sec3:conserved energy current}
			\begin{aligned}
				J_{\nu,c}=&\sum_{j=1,2}\left[  \partial_\nu\phi_j\partial_t\phi_j -\frac{1}{2}\mu^2g_{\nu0}\phi_j^2
				-\frac{1}{2}g_{\nu 0}g^{\rho\sigma}\left( \partial_\rho\phi_j\partial_\sigma \phi_j+q^2A_\rho A_\sigma\phi_j^2 \right)  \right].
			\end{aligned}
		\end{equation}
		which shares the same form as the RN black hole \cite{Laurent_Di_Menza_2015}. 
		Through the energy current, we can also construct the energy flux at a fixed radial distance, by integrating the $\nu=1$ component of Eq. \eqref{sec3:conserved energy current}, 
		\begin{equation}\label{sec3:energy flux}
			\mathcal{F}_r
			=\int (r^2+\lambda^2)~\diff t\diff \theta \diff \varphi~ \text{Re}{(\partial_t \Phi \partial_x \Phi^*)}.
		\end{equation} 
		Substituting Eq. \eqref{sec3:seperation of variables} into Eq. \eqref{sec3:energy flux}, we obtain 
		\begin{equation}\label{flux2}
			\mathcal{F}_r=\frac12 \int \diff t~ (\partial_t\psi \partial_x \psi^*+\partial_t\psi^* \partial_x \psi),
		\end{equation}
		where the normalization of spherical harmonic functions $ \int_0^\pi \diff \theta \int_0^{2\pi}\diff \varphi \sin\theta Y_{lm}(\theta,\varphi) Y_{lm}^*(\theta,\varphi) =1$ is used. The Eqs. \eqref{sec3:energy flux} and \eqref{flux2} measure the energy that crosses a hypersurface where $x$ is a constant. Suppose we place this hypersurface near the event horizon,
		since only the incident waves that fall into the black hole are involved, we can calculate the negative energy passing through, and equivalently we obtain the energy gain of  the scalar field according to the energy conservation.

		\subsection{Numerical methods and boundary conditions}
		Here we present the numerical strategy to solve the partial differential equation \eqref{sec3:time-domain eom} in the time domain, where we adopt the finite difference method combined with the method of lines.
		First, we uniformly discretize the spatial coordinates with a grid spacing of $\delta x$. Therefore, the spatial derivatives after discretization can be expressed as
		\begin{align}
			&\partial_x \psi=\frac{\psi_{i+1}-\psi_{i-1}}{2\delta x}, \\
			&\partial_x^2\psi=\frac{\psi_{i+1}-2\psi_{i}+\psi_{i-1}}{\delta x^2},
		\end{align}
		where $\psi_i$ denotes $\psi(t,x_m-i\times \delta x)$ and $x_m$ is the mirror's position. After discretizing the spatial coordinates, we obtain a system of ordinary differential equations (ODEs) containing only time derivatives. We employ the fourth-order Runge-Kutta method to numerically integrate this system of ODEs, where we set $\delta x/M=0.2$ in the computation.
		
		For a black hole bomb, since all waves at the mirror are completely reflected back, the boundary condition at the mirror surface can be simply represented as a Dirichlet boundary condition, namely that the value of the solution is set to vanish, corresponding to $\psi(t,x_m)=0$. Another boundary condition occurs at the horizon, however	
		since the tortoise coordinate spans from $ -\infty $ to $ x_m $, truncating the computational domain at the left becomes necessary. Due to the fact that applying Dirichlet or Neumann boundary conditions directly at the truncated boundaries introduces artificial wave reflections, we employ the technique of Perfectly Matched Layers (PML) boundary \cite{Antoine18082017} to suppress outgoing waves effectively, in which 
		an absorbing region at the computational boundary is introduced, so that outgoing waves are attenuated through damping profiles. 
		
		To ensure the simplicity of the equation after applying the PML, we first change the variable to $ \psi=\me ^{\mi(qA_0+\sqrt{W_\text{eff}})t} u(t,x) $, then we obtain
		\begin{equation}
			\partial_t^2u-\partial_x^2u+2\mi\sqrt{W_\text{eff}}\partial_tu=0.
		\end{equation}
		In the frequency domain, the time derivative is replaced by $-i\omega$. By further applying the substitution $ \partial_x^2\to \frac{1}{1+\mi\sigma(x)/\omega} \partial_x^2 $, we obtain the modified partial differential equation compatible with the PML framework in the frequency domain
		\begin{equation}\label{sec3:equation frenquency domain}
			\begin{aligned}
				&(-\mi\omega)^2u-\partial_x^2u+2\mi\sqrt{W_\text{eff}}(-\mi\omega)u
				+\sigma(x)(-\mi\omega)u +2\mi\sigma(x) \sqrt{W_\text{eff}}u=0,
			\end{aligned}
		\end{equation}
		where $\sigma(x)$ is defined as continuous function that vanishes in the physical domain outside the absorbing layer and assumes non-zero values only within the absorbing region. To eliminate wave reflections at the interface between the physical domain and the absorbing layer, $\sigma(x)$  must be carefully formulated as a smoothly varying function to guarantee effective wave attenuation. Transforming the above equation \eqref{sec3:equation frenquency domain} from the frequency domain back to the time domain, the equation becomes
		\begin{equation}
			\begin{aligned}
				&\partial_t^2u-\partial_x^2u+\sigma\partial_t u+2\mi\sqrt{W_\text{eff}}\partial_t u 
				+2\mi\sigma \sqrt{W_\text{eff}}u=0.
			\end{aligned}
		\end{equation}
		Changing the variables to $\psi$, we get
		\begin{equation}\label{sec3:PML equation}
			\begin{aligned}
				&\partial_t^2\psi-\partial_x^2\psi-2\mi qA_0\partial_t\psi+V_\text{eff}\psi+\sigma\partial_t\psi
				+\mi\sigma(\sqrt{W_\text{eff}}-qA_0)\psi=0.
			\end{aligned}
		\end{equation}
		When $\sigma(x)=0$, Eq.\eqref{sec3:PML equation} reduces to the original partial differential equation \eqref{sec3:time-domain eom}, demonstrating that the presence of $\sigma(x)$ preserves the intrinsic wave dynamics within the physical domain. Conversely, in regions where $\sigma \ne 0$ (within the absorbing layer), the modified equation \eqref{sec3:PML equation} incorporates three dissipative terms, leading to exponential attenuation of wave amplitudes until complete extinction occurs. This ensures that outgoing waves are absorbed without contaminating the physical domain. Note that our substitution scheme modifies the second spatial derivative as $\partial_x^2\to \frac{1}{1+\mi\sigma(x)/\omega}\partial_x^2$, rather than the conventional first-order derivative replacement $\partial_x\to \frac{1}{1+\mi\sigma(x)/\omega}\partial_x$. This approach leverages the combined dissipative and damping terms to suppress reflections when a sufficiently thick absorption layer is implemented. In contrast, adopting the standard substitution would introduce prohibitive complexity into the resulting modified partial differential equation.

		\begin{figure}[htbp]
			\centering
			\includegraphics[scale=0.35]{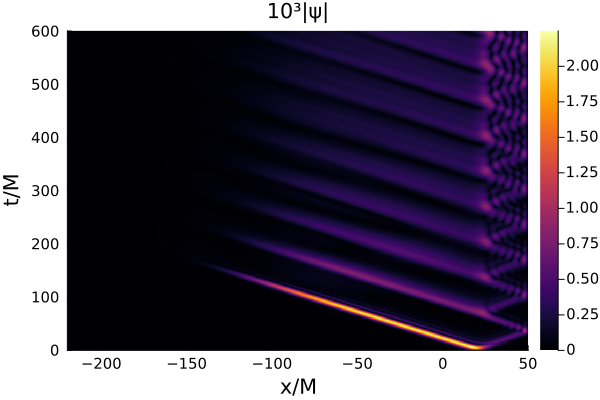}
			\includegraphics[scale=0.35]{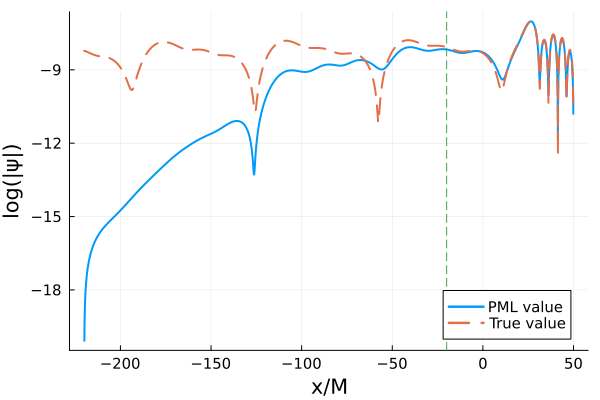}
			\caption{The validation of the effectiveness of PML boundary condition. The left panel illustrates the continuous reflection and transmission of the scalar field between the mirror and the black hole. The right panel presents a comparison between the scalar field amplitude after applying PML boundary conditions and the true amplitude. The dash green line is the boundary of the PML layer. }
			\label{sec3:pml boundary}
		\end{figure}
		
		Next, we evaluate the efficacy of PML boundary conditions. This verification involves two key aspects: first, ensuring that no reflected waves are generated either at the boundary or within the absorbing layer itself; second, confirming whether the absorbing layer fully attenuates the incoming waves. We set the truncation boundary at $x_0=-20M$ with a PML layer width of $200M$. Additionally, to evolve the scalar field, we need to choose a proper initial condition for the scalar field,
		which does not affect the growth rate of the scalar field \cite{Fu:2025ztk}. The initial condition is 
		chosen
		as a Gaussian wave packet for simplicity
		\begin{equation}\label{sec3:initial condition}
			\1\{\begin{aligned}
				&\psi(t=0,x)=\psi_0~ \me^{-(x-x_\text{in})^2/\alpha^2}, \\ & \partial_t \psi(t=0,x)=0,
			\end{aligned}\2.
		\end{equation}
		with $\psi_0$ the peak amplitude of the wave packet, $x_\text{in}$ its initial center position, and $\alpha$ the spatial width of the wave packet. 
		We set $\psi_0=10^{-3}$, $\alpha=5M$ to include as many frequency components as possible and choose the initial position to be $ x_\text{in}=20M$, where the effective potential is sharpest, as shown in Fig.\ref{sec3:potential}. 
		We present the validation results in Fig.\ref{sec3:pml boundary}. The left panel demonstrates that the incident wave undergoes transmission and reflection at the initial place, with the par incident into the absorbing layer gradually diminish until completely disappearing, while the other part on the right side are fully reflected back at the mirror.
		The right panel shows the wave amplitude at $ t=600M $, where it can be observed that the application of the PML boundary condition does not affect the values within the physical domain.
		
		\subsection{Evolution of the scalar field and impact of the regularization parameter}
		We proceed to study the evolution of the scalar field in the first two cases mentioned in \cite{Di_Menza_2020}: Type I, 
		the mirror's position is set 
		outside the effective ergosphere; Type II, the mirror's position is set  inside the effective ergosphere.
		When extracting the eigenfrequencies of scalar field, we employ the fast Fourier transform (FFT) to obtain the real parts and utilize digital filter method to extract the imaginary parts of the eigenfrequencies. To measure the energy gain of the scalar field, we adopt Eq. \eqref{sec3:energy flux} to numerically calculate it. 
		
\begin{figure}[htbp!]
			\centering
			\includegraphics[scale=0.25]{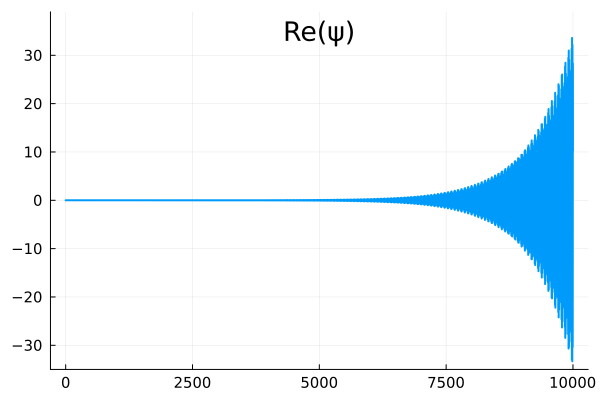}
			\includegraphics[scale=0.25]{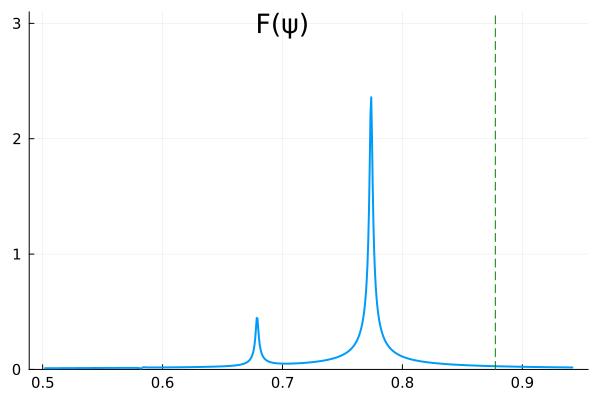}
			\includegraphics[scale=0.25]{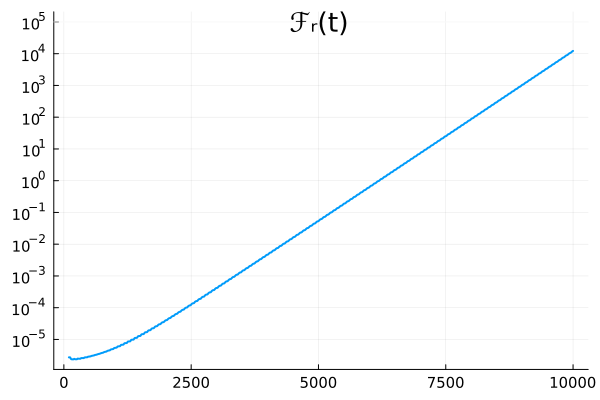}
			\includegraphics[scale=0.25]{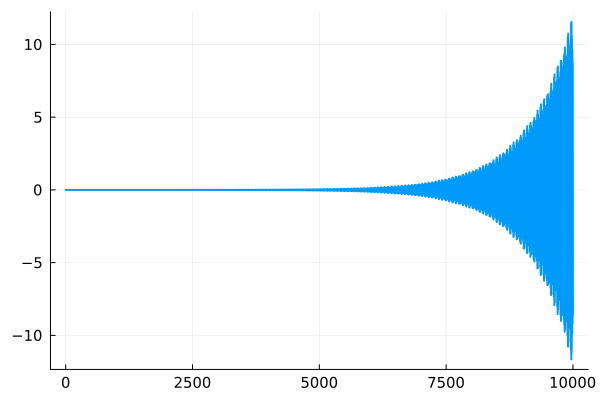}
			\includegraphics[scale=0.25]{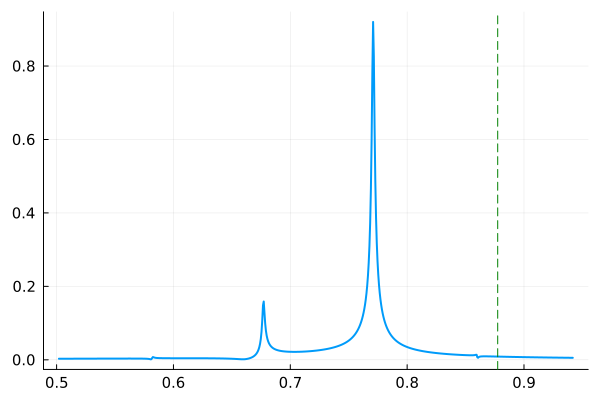}
			\includegraphics[scale=0.25]{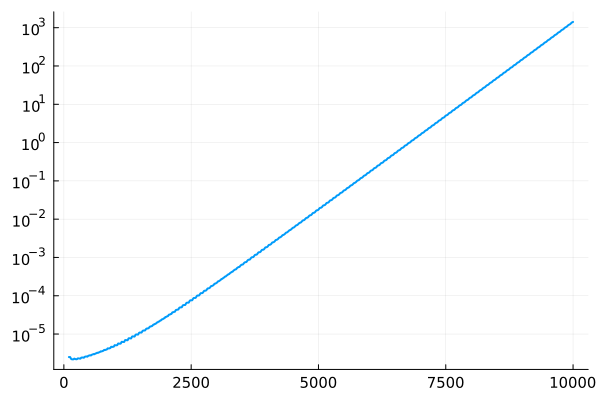}
			\includegraphics[scale=0.25]{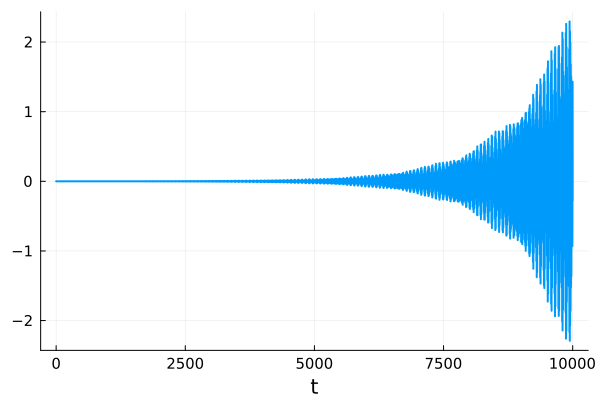}
			\includegraphics[scale=0.25]{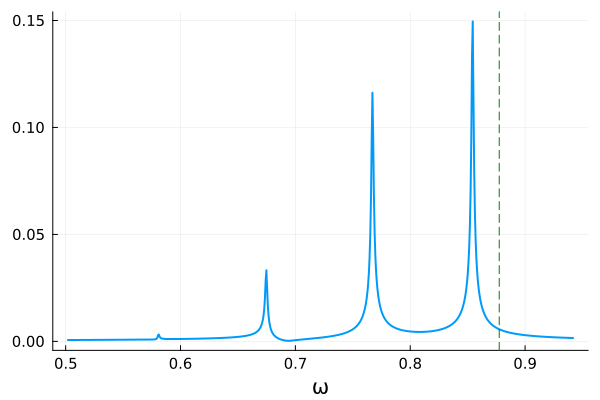}
			\includegraphics[scale=0.25]{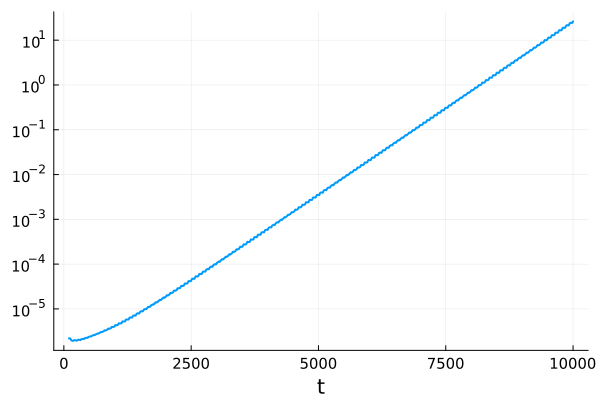}
			\caption{Evolution of the scalar field in a Type I BH bomb. The position of the mirror is at $x_m=50M$, and the parameters are set as $ Q/M=0.9 $, $ \mu M=0.3 $, $ qM=1.4 $, $l=0$. From left to right, the three columns represent the field amplitude, power spectrum and energy flux (measured at $x_s=-40M$) . From top to bottom, the values of $\lambda/M$ are 0.0, 0.8 and 1.2 respectively.  The green dash line represents the critical superradiant frequency which is approximately $\o_c M\simeq 0.8775$. 
			}
			\label{sec3:fig1}
		\end{figure}

		In a Type I BH bomb,  
		we position the mirror outside the ergosphere at $x_m=50M$, set the absorbing layer to span from $x=-50M$ to $x=-250M$, and evolve the system to $t_\text{end} = 10^4M$. The parameters in effective potential are taken as 
		$ Q/M=0.9$, $ \mu M=0.3 $, $ q M=1.4 $ and $l=0$, with the critical superradiant frequency $\omega_c = qA_0(r_+) =0.8775/M$. The initial condition is set by a Gaussian wave packet described by Eq. \eqref{sec3:initial condition}, with $\psi_0=10^{-3}$, $x_\text{in}=20M$ and $\alpha=5M$. Additionally, the energy flux at $x_s=-40M$ is measured and the power spectrum is computed at the end of the time domain.
		The results are shown in the Fig.\ref{sec3:fig1}, with three columns from the left to right, representing the real part of amplitude, power spectral and energy flux, respectively. From top to bottom panels, the corresponding values of $\lambda/M$  are taken as 0.0,  
		0.8 and 1.2 respectively. Owing to the constraints imposed by the mirror, not all frequencies are sustained; consequently, the power spectrum exhibits multiple peaks that correspond to the eigenfrequencies, and
		the eigenfrequencies are listed in Table \ref{sec3:table1}. 
		It is observed that the wave amplitude, the real and imaginary parts of the eigenfrequencies and energy flux simultaneously decrease with the increase of $\lambda$. Especially, 
		as shown in the bottom plots in Fig. \ref{sec3:fig1}, the real part of the fourth eigenfrequency decreases sufficiently to meet the superradiant condition, resulting in a new distinct growing mode at this configuration, 
		though the amplification of the scalar field remains diminished due to the reduced imaginary part of the eigenfrequencies.
		
\begin{figure}[h]
			\includegraphics[scale=0.3]{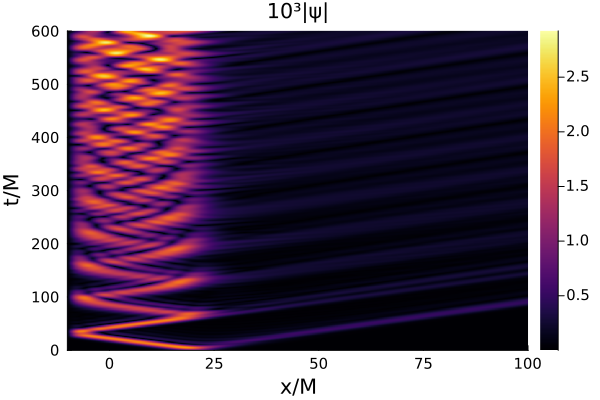}
			\includegraphics[scale=0.3]{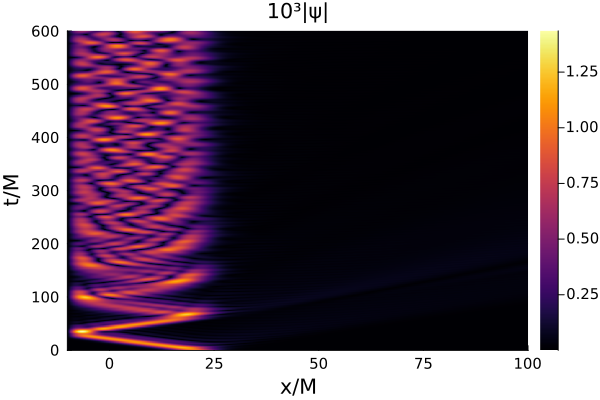}
			\caption{Spacetime diagram of the field strength evolution in a Type II BH bomb, with the mirror setting at $x_m=-10M$, and the PML ranging from $100M-300M$. The sub-case of light particles ($\m<\o_c$) is shown on the left, with a parameter choice of $ \mu M=0.3$, $ Q/M=0.9$, $ q M=1.0 $ and $\omega_cM =0.8775$; while the sub-case of heavy particles ($\m>\o_c$) is shown on the right, with the choice of $\mu M=0.6$, $Q/M=0.9$, $qM=0.8$ and $ \omega_c=0.5014$.
			}
			\label{sec3:aaa}
		\end{figure}

		In a Type II BH bomb, the scalar field undergoes reflection and transmission between the mirror and the boundary of the ergo-region. We place the mirror within the ergosphere at $x_m=-10M$, other parameters chosen as the same as in Type I, while we measure the energy flux at $x_s=70M$ in this case.
		Yet, there are still two situations in this scenario,  
		when the superradiance frequency exceeds the scalar field mass $\o_c>\m$, 
		there could exist the eigenfrequencies $\o$ lies between them, allowing
		the field to propagate outward through the boundary of the ergoregion; 
		conversely, when the field's mass exceeds the superradiant eigenfrequency $\m>\o_c$, there's also a chance for eigenfrequencies that satisfy $\o<\o_c$, rendering the field confined within the ergoregion, 
		to form a persistent scalar cloud surrounding the black hole. The evolutions of the scalar field in these two subcases are shown in Fig.\ref{sec3:aaa}, where one can clearly see that the waves appear outside the effective potential when $\o_c>\o>\m$ and disappear when $\m>\o_c>\o$.

\begin{figure}[htbp!]
			\centering
			\includegraphics[scale=0.25]{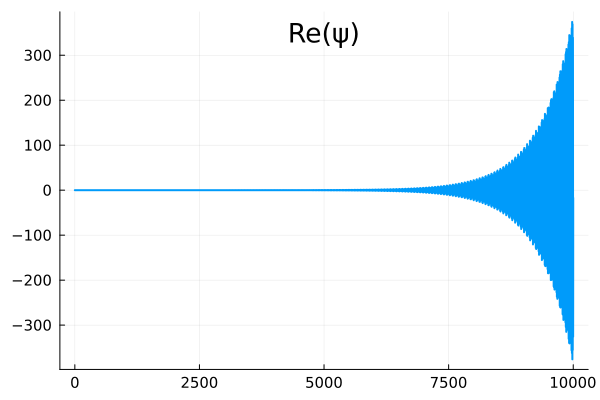}
			\includegraphics[scale=0.25]{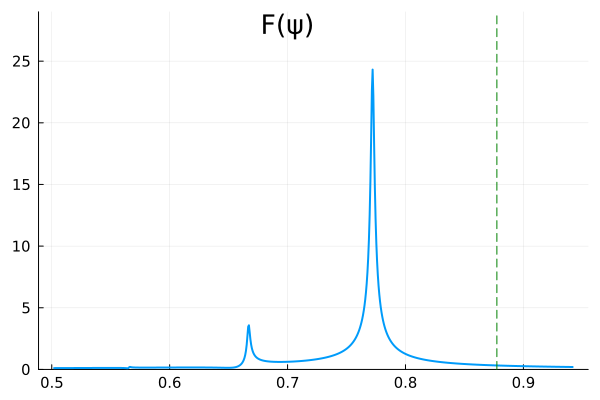}
			\includegraphics[scale=0.25]{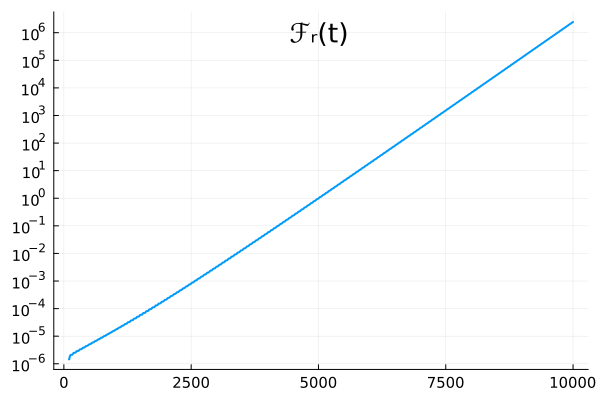}
			\includegraphics[scale=0.25]{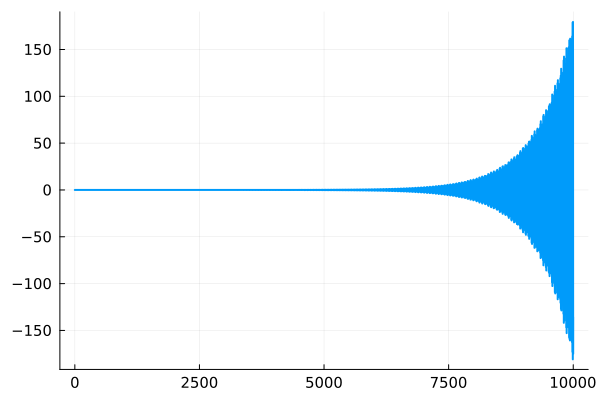}
			\includegraphics[scale=0.25]{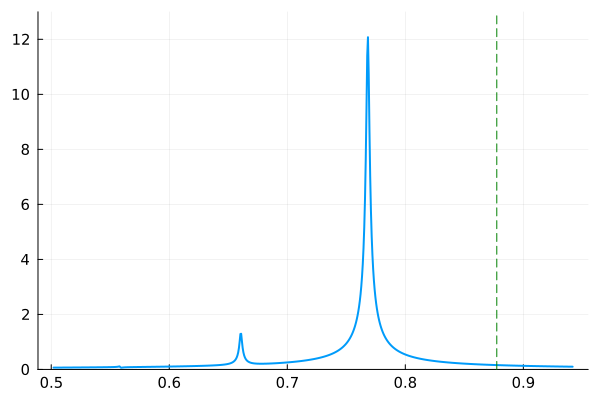}
			\includegraphics[scale=0.25]{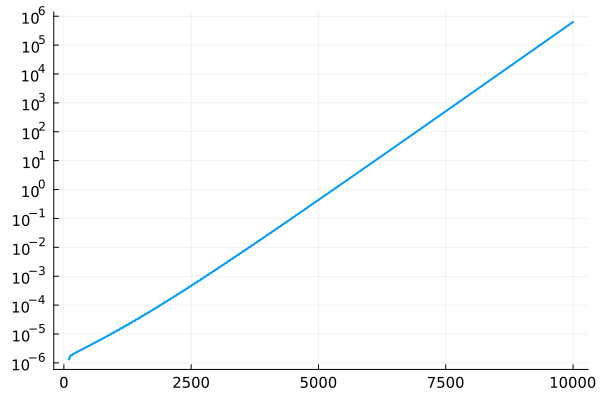}
			
			\includegraphics[scale=0.25]{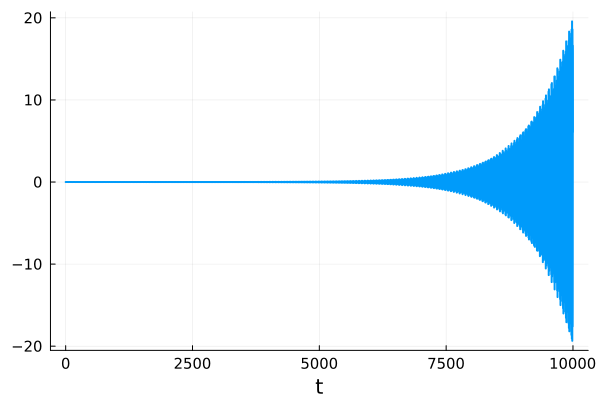}
			\includegraphics[scale=0.25]{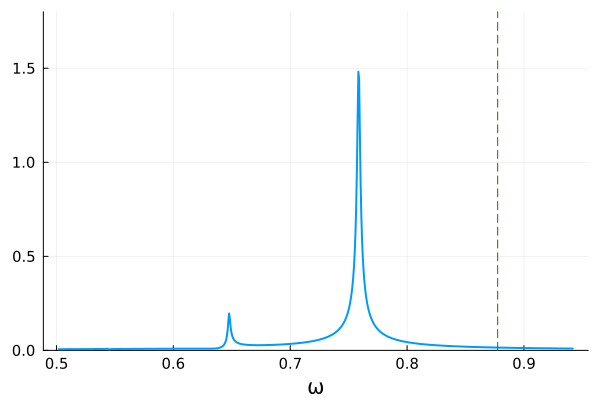}
			\includegraphics[scale=0.25]{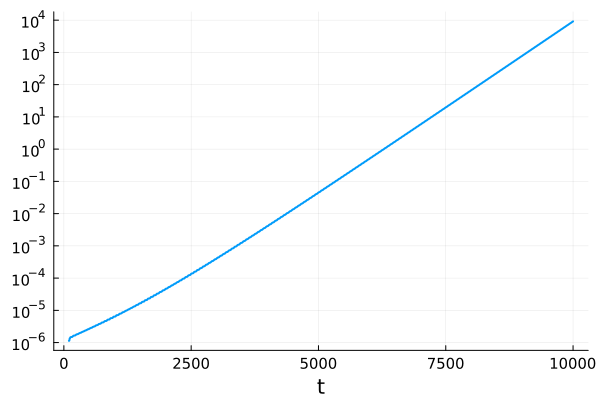}
			
			\caption{Evolution of the scalar field in a Type II BH bomb, in the sub-case of light particles $\m<\o_c$. The position of the mirror is at $x_m=-10M$, and the parameters are set as $ Q/M=0.9 $, $ \mu M=0.3 $, $ qM=1.4 $, $l=0$. From left to right, the three columns represent the field amplitude, power spectrum and energy flux (measured at $x_s=70M$) . From top to bottom, the values of $\lambda/M$ are 0.0, 0.8 and 1.2 respectively.  The green dash line represents the critical superradiant frequency which is approximately $\o_c M\simeq 0.8775$. }
		\label{sec3:fig2}
	\end{figure}

		\begin{table}[htbp!]
			\centering
			\begin{tabular}{|c|c|c|c|c|}
				\hline
				$\lambda/M$ & $\omega_0 M$ &  $\omega_1 M$ & $\omega_2 M$ & $\omega_3 M$ \\ \hline
				0.0 & $0.5831+\mi8.073\times 10^{-4} $ & $0.6785+\mi1.125\times 10^{-3} $ & $0.7740+\mi1.243\times 10^{-3} $ & \slash \\ 
				0.8 & $0.5818+\mi7.319\times 10^{-4} $ & $0.6773+\mi1.014\times 10^{-3} $ & $0.7709+\mi1.138\times 10^{-3} $  & \slash\\ 
				1.2 & $0.5811+\mi6.243\times 10^{-4} $ & $0.6747+\mi8.388\times 10^{-4} $ & $0.7671+\mi9.094\times 10^{-4} $  & $ 0.8544 + \mi 9.832\times 10^{-4} $\\ \hline
			\end{tabular}
			\caption{Numerics of eigenfrequencies in a Type I BH bomb, corresponding to the middle panel of
				Fig. \ref{sec3:fig1}, both the real and imaginary parts are presented. The table is divided with respect to different choices of the quantum parameter $\l/M$. Three modes appear in the $\l/M=0, 0.8$ case, while a new eigen-mode shows up in the $\l/M=1.2$ case.
			}
			\label{sec3:table1}
		\end{table}
		\newpage
		For $\o_c>\m$, we also present the field evolution
		in Fig.\ref{sec3:fig2}. Again, as $\lambda$ increases, both the scalar field amplitudes and energy fluxes exhibit monotonic decrease, and so do the imaginary parts of the eigenfrequencies. Numerics corresponding to the eigenvalues are shown in Table \ref{sec3:table2}, where there are only two eigen-modes and no new modes appearing in the $\l/M=1.2$ case, different from that in Type I BH bomb.
		To elucidate this phenomenon, analytical expressions for the eigenfrequencies are to be derived, particularly for cases where eigenfrequencies lie near the superradiance critical frequency. However, due to the introduction of the parameter $\lambda$, obtaining such analytical solutions proves highly challenging at present.
		  The emergence of a new eigenfrequency satisfying the superradiance condition in the Type I BH bomb scenario is relatively uncommon. 
		A possible qualitative explanation is that under identical parameter values, the absence of a left-side mirror in Type I results in a deeper potential well, enabling it to support a greater number of eigenstates. In contrast, the presence of the left mirror in Type II imposes a constraint shallowing the effective potential well, thereby reducing its capacity to accommodate eigenstates. We leave the physical interpretation for this difference in the future study.

\begin{figure}[h!]
		\centering
		\includegraphics[scale=0.3]{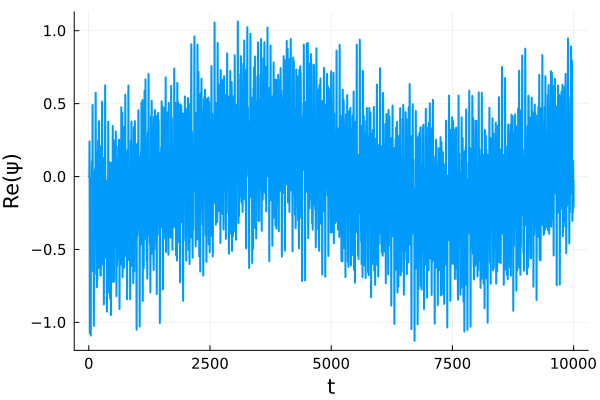}
		\includegraphics[scale=0.3]{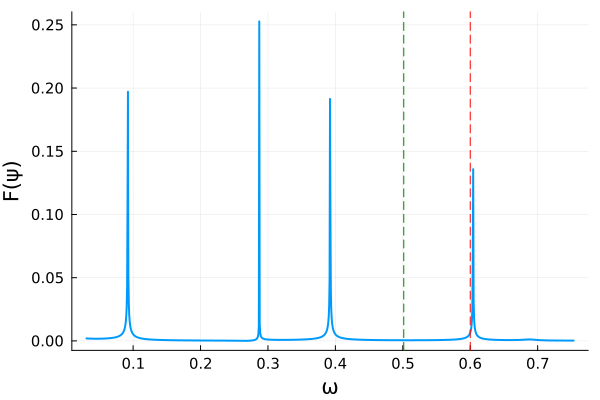}
		\includegraphics[scale=0.3]{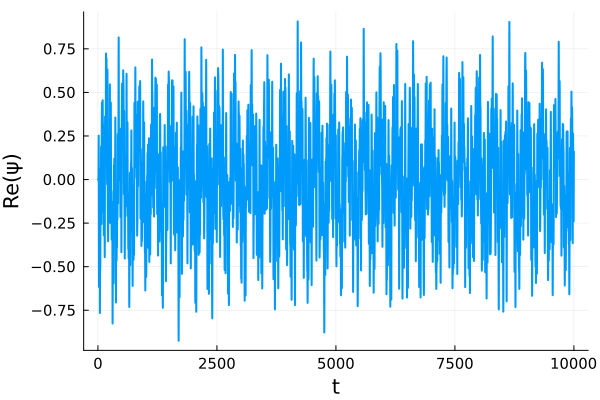}
		\includegraphics[scale=0.3]{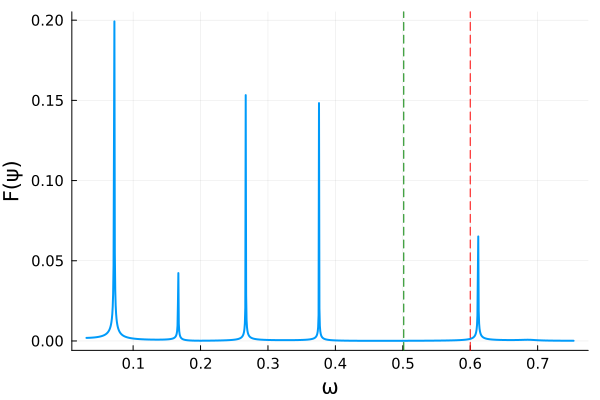}
		\caption{Evolution of the scalar field in the sub-case of $\m>\o_c$ in a Type II BH bomb. The wave amplitude evolution measured at $x=0$ is shown on the left and the corresponding power spectrum is shown on the right, with $\m M=0.6$ denoted by the red dashed line and $\o_c\simeq0.5$ by the green dashed line. Other parameters are chosen as $Q/M=0.9$, $qM=0.8$, $\l/M=$ 0.0 (Top) and 1.2 (Bottom). 
		}
		\label{sec3:mu}
	\end{figure}

		For $\m>\o_c$, we set $\mu M=0.6$, $Q/M=0.9$, $qM=0.8$, and the value of $\l$ is taken as 0.0 and 1.2, thus the corresponding superradiant frequency is $ \omega_c=0.5014$. The evolutions are shown in Fig.\ref{sec3:mu}, where wave amplifications are absent, yet eigen-frequencies are drawn in the right panel. Compared with the case where $\lambda = 0.0$, when $\lambda = 1.2$, there exists a new eigenfrequency within the range of $\omega < \omega_c$. It should be noted that, unlike the one found in the Type I case, this new eigenfrequency is not superradiantly amplified.

	\begin{table}[h!]
		\centering
		\begin{tabular}{|c|c|c|}
			\hline
			$\lambda/M$ & $\omega_0M$ & $ \omega_1M$ \\ \hline
			0.0 & $ 0.6672+ \mi 1.234\times 10^{-3} $ & $ 0.7721+\mi1.485\times10^{-3} $ \\ 
			0.8 & $ 0.6609 + \mi 1.095\times10^{-3} $ & $ 0.7683+\mi1.428\times10^{-3} $  \\
			1.2 & $ 0.6477+\mi 0.929\times10^{-3} $ & $0.7583 + \mi 1.235\times 10^{-3} $ \\ \hline
		\end{tabular}
		\caption{
			{
				Numerics of eigenfrequencies in a Type II BH bomb, corresponding to the middle panel of Fig. \ref{sec3:fig2}, both the real and imaginary parts are presented. The table is divided with respect to different choices of the quantum parameter $ \l/M $. In this light-particle sub-case, two distinct modes emerge across the three specified values of $\lambda/M$.}
		}
		\label{sec3:table2}
	\end{table}

	~\\

	\section{Conclusion}\label{sec3}
	In this work, we studied
	the superradiance 
	for a charged scalar field propagating in the spacetime of a black-bounce-RN black hole, which could be obtained by
	the replacement $ r\rightarrow \sqrt{r^2+\lambda^2} $ in the RN metric to reduce the curvature singularity.
	The existence of the quantum parameter $ \lambda$ shallowed the effective potential in the Regge-Wheeler equation hence weakened the negativity of the energy a field/particle can get, which further reduced the superradiant abilities both in the scattering experiment and the BH bomb phenomenon. We verified the effect in these two scenarios numerically and found several new features that may provide insights to related research.
	
	
	In the superradiant scattering experiment, we focused on the influence that parameters had on the amplification factor defined as $Z=\left|{\mathcal{R}}\right|^2/\left|{\mathcal{I}}\right|^2-1$. As expected, 
	numerical results reveal that the quantum parameter $\lambda$ indeed reduced the amplification factor over the whole 
	frequency range, which indicates that black-bounce-RN BH has a weaker superradiant amplification than the usual RN BH.
	In a similar manner, the larger field mass suppresses the amplification in all frequencies, in agreement with the physical intuition that heavier particles are dominated more by the gravitational attraction and harder to decay/transit into negative energy states. Another pair of parameters, the BH and the field charge, affect the amplification factor in a coherent way, they raised the critical frequency as the superradiant condition indicates, and notably enhance it at high frequencies while suppress it in low frequencies, which are consistent with the behaviors observed in the RN BH case as they are independent of the quantum parameter. 
	
	
	In the black hole bomb phenomenon, we investigated in the time domain the evolution of the charged scalar field in a fixed background of black-bounce RN BH,
	subjected to the mirror-like boundary condition. In two types of BH bombs, where the mirror is placed outside or inside the ergoregion, 
	the amplitude evolution, the eigenfrequencies and the energy growth rate are simulated. 
	Technically, we employed the finite difference method combined with the method of lines to numerically solve the differential equation for the field amplitude evolution. The Dirichlet boundary condition is chosen at the mirror to fully reflect the wave, while the purely ingoing boundary condition at the event horizon is implemented by the Perfectly Matched Layers, whose validity was tested in our numerical simulations. The eigenfrequencies  were extracted by the discrete Fourier transform and digital filter method. To measure the energy flux,
	we constructed analytically
	a conserved energy current based on the stress-energy tensor of the scalar field, and invoke the numerical values obtained to achieve it.
	%
	Again as expected, the wave amplification and the energy growth rate are weakened by a larger value of the quantum parameter $\l$. However, we discovered that in the Type I BH bomb case when $\l/M=1.2$ a new distinct eigen-mode appeared while in the light-particle sub-case of the Type II BH bomb there is no such an extra mode, compared to those of $\l/M=0, 0.8$. The mathematical observation attributes this to a decrease of the real part of the frequency, but the physical origin deserves a further analytical study in the future. Moreover, in the case of Type II bomb, we investigated further when the field mass is larger than the critical frequency $\m>\o_c$, and found that a new eigenfrequency emerges under the $\lambda=1.2$ condition when compared to the low $\l$ values; however, unlike the Type I case, it is not superradiantly amplified.
	
	The study is straightforward to be extended to several directions. First, in the current paper, we only focus on the influence of quantum parameter in the spherically symmetric charged black-bounce black hole. In contrast to it, the effects of quantum parameters in nonsingular rotating black hole configurations have yet to be systematically investigated. This will advance our understanding of how quantum parameter governs superradiant energy extraction processes and black hole bomb instability. Moreover, the present study employs an artificially engineered reflecting boundary condition beyond the event horizon to constrain the scalar field. Nevertheless, the implementation through nonlinear interaction terms within the scalar field framework provides an alternative pathway to achieve analogous confinement effects \cite{Zhang:2023qtn}. In addition, this investigation confines its scope to evolution of scalar field, while further studies are needed to explore the dynamics of black hole mass and charge.
	

	\section*{Acknowledgements}
	The work of Xiangdong Zhang is supported by the National Natural Science Foundation of China with Grant No. 12275087. The work of Gui-Rong Liang is supported by the National Natural Science Foundation of China with Grant No. 12175099 and Jiangsu Funding Program for Excellent Postdoctoral Talent with No. 2024ZB752.

	\bibliographystyle{JHEP}
	\normalem
	\bibliography{ref}
	
\end{document}